\documentclass[12pt]{iopart}

\usepackage{iopams}
\usepackage{graphics}
\usepackage[next]{inputenc}
\usepackage[dvips]{epsfig}

\begin{document}

\title[Thermodynamics of XXZ chain around the factorizing field]
{Thermodynamic behavior of the XXZ Heisenebrg $s=1/2$ chain around the factorizing magnetic field}

\author{J. Abouie$^{1,3}$, A. Langari$^2$ and M. Siahatgar$^2$}
\address{$^1$ Department of physics, Shahrood University of Technology, Shahrood 36199-95161,
Iran\\
$^2$ Department of Physics, Sharif University of Technology,
Tehran 11155-9161, Iran\\
$^3$ School of physics, Institute for Research in Fundamental Sciences (IPM),
Tehran 19395-5531, Iran}

\ead{jahan@shahroodut.ac.ir}

\date{\today}

\begin{abstract}
We have investigated the zero and finite temperature behaviors of the anisotropic
antiferromagnetic Heisenberg XXZ spin-1/2 chain in the presence of a transverse
magnetic field ($h$).
The attention is concentrated on an interval of magnetic field between
the factorizing field ($h_f$) and the critical one ($h_c$).
The model presents a spin-flop phase for $0<h<h_f$ with an energy scale which
is defined by the long range antiferromagnetic order while it undergoes an
entanglement phase transition at $h=h_f$.
The entanglement estimators clearly show that the entanglement is lost
exactly at $h=h_f$ which justifies different quantum correlations on both
sides of the factorizing field.
As a consequence of zero entanglement (at $h=h_f$) the ground state is known exactly
as a product of single particle states which is the starting point for initiating a spin wave
theory.
The linear spin wave theory is implemented to obtain the specific heat and thermal
entanglement of the model in the interested region.
A double peak structure is found in the specific heat around $h=h_f$ which manifests
the existence of two energy scales in the system as a result of two competing orders
before the critical point.
These results are confirmed
by the low temperature Lanczos data which we have computed.
\end{abstract}


\maketitle

\section{Introduction}

The zero temperature phase diagram (i.e. quantum phase diagram) of a
model gives important information on the low temperature behaviors
of the system \cite{Sach 99, Vojt 03}. The anisotropic antiferromagnetic Heisenberg
(XXZ) spin-1/2 model shows different quantum phases with respect to
a symmetry-breaking transverse field (non-commuting magnetic
field) \cite{dmitriev, Lang04, Lang2-06}. The non-commuting field imposes
quantum fluctuations into the ground state which can induce new
phases.
$Cs_2CoCl_4$ is a quasi-one dimensional
spin-1/2 XY-like antiferromagnet with weak inter-chain
couplings ($J^{\prime}/J=0.014$) which can be studied in terms of XXZ chain with anisotropy parameter
$\Delta=0.25$ \cite{Kenz02, Radu 02}. The scaling behavior and
quantum phase diagram of the XXZ model in the presence of a transverse
magnetic field ($h^x$) have been investigated \cite{dmitriev, Lang04, Lang2-06, caux}.

Recently, the quantum spin models received much attentions from the quantum
information point of views. These are prototype models to implement and examine
the idea of quantum computations which requires the quantum correlations
measured by entanglement. Hence, disentangled ground states
have to be avoided for such implementations. On the other hand, the
zero entanglement property of a ground state provides an exact form for it
in terms of product of single particle states. The investigation in this direction
resumed recently following the seminal work of J. Kurman and his collaboratores\cite{Kurm82}.
Several efforts have been devoted to this direction which is
important for condensed matter researchers, i.e finding an exact (factorized) ground
state even at particular values of the coupling constants \cite{Rosc04,Giam08,Giam09,Reza09}.
The factorized (exact) ground state is an accurate starting point
to investigate the quantum nature of a phase close to the factorizing point
in addition to some exact knowledge which gives at the factorized point.
This property is implemented in this article to initiate a spin wave theory
to describe the thermodynamic properties of the XXZ model in the presence of
a transverse magnetic field.

The U(1) symmetry of the XXZ model is lost upon adding the transverse
magnetic field.
Initially, a perpendicular antiferromagnetic order is stabilized by promoting a
spin-flop phase (which has a partial moment projection along the field direction).
At the factorizing field ($h_f$) in the spin-flop phase
the ground state is known exactly
as a direct product of single spin states and the
staggered magnetization along  $y$-direction is close to its maximum
value. In our model the factorizing field is $h_f=J \sqrt{2(1+\Delta)}$, where quantum fluctuations
are uncorrelated and the ground state is the classical one
($J$ is the scale of energy and $\Delta$ is the anisotropy parameter). 
For slightly larger magnetic field very close to the critical one ($h_f<h<h_c$) the antiferromagnetic order becomes unstable and the staggered
magnetization falls rapidly to vanish at the critical point.
For $h>h_c$ the spins become aligned in $x$ direction and a
fully polarized phase will be appeared. The region between the
factorizing and the critical fields ($h_f<h<h_c$) is the main issue
of our study which is not well understood so far.

The zero temperature properties of the intermediate region ($h_f<h<h_c$)
induces its signature into the thermodynamic functions of the model.
We have found that this region is characterized by two energy scales and
its fingerprint will appear as a double peak in the specific heat.
Moreover, the existence of more than an energy scale in the model can
be related to a spontaneous symmetry breaking (SSB) \cite{Rosc04}. Each
broken phase is described by an order parameter which can be zero in
the disordered phase. For the aforementioned model the symmetry breaking
occurs where the staggered magnetization becomes zero.

The thermodynamic properties of the spin 1/2 XXZ
chain in a transverse magnetic field have been studied using the Low
temperature  Lanczos method \cite{Siah08}.
However, in this paper we will focus on the intermediate
values of $h$ and more precisely on the region where a double peak appears
in the specific heat of the model.
We implement the exact factorized ground state at $h=h_f$ (where the entanglement
vanishes) to build up a spin wave theory appropriate to describe the model in the
intermediate values of the magnetic field. In the linear spin wave approximation we
calculate the specific heat and thermal entanglement of the model. Moreover, the
spin wave theory gives two different types of quasi particles which are representing
the two energy scales. We have also studied both the zero and finite temperature
properties of the model on a finite chain using the low temperature Lanczos method \cite{aichhorn}.
Our numerical results are in agreement with the spin wave theory counterparts.
In the next section, we will briefly review the zero temperature properties of the
model from the quantum information point of view. We will provide the spin wave theory
in Sec.\ref{spwt} where the quantum property, specific heat and thermal entanglement
are obtained. Finally, the numerical Lanczos results will be presented
in addition to discussions on the mentioned topics.

\section{Zero temperature phase diagram}

\begin{figure}[h]
\hspace{4 cm}
\includegraphics[width=10cm, angle=0]{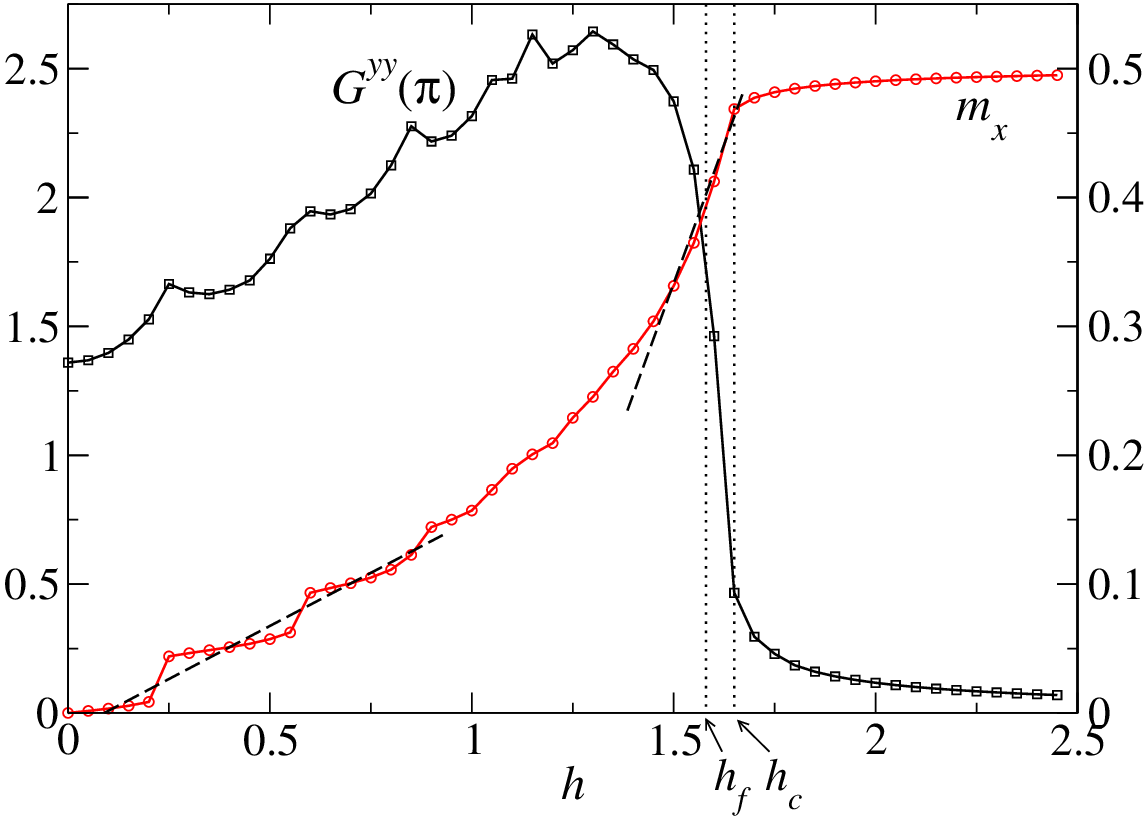}
\caption{\label{magnet} (color online) Results of numerical zero temperature Lanczos method for the magnetization along  $x$-direction (right vertical axis) and the $y$-component spin structure factor at momentum $\pi$ (left vertical axis) versus the
transverse magnetic field $h$ for a chain with $N=20$ and
$\Delta$=0.25. The magnetization in $z$-direction is zero.} \label{Mx}
\end{figure}

The anisotropic spin-1/2 Heisenberg model in the presence of a transverse magnetic
field is described by the following Hamiltonian,
\begin{equation}
H=J \sum_i(S_i^xS_{i+1}^x+S_i^yS_{i+1}^y+\Delta
S_i^zS_{i+1}^z+h S_i^x),
\end{equation}
where $S_i^{\alpha}$'s are spin-1/2 operators, $\Delta$ is the anisotropy
parameter, $h$ is proportional to the transverse magnetic field
and the exchange antiferromagnetic coupling $J$ which defines the scale of energy is set to one.
This model has several quantum phases with respect to
the transverse magnetic field and anisotropy parameter ($\Delta$).
The $\alpha$-component spin structure factor at momentum $p$ is defined by
\begin{equation}
G^{\alpha \alpha}(p)=\sum_{x=1}^N \langle S_1^{\alpha} S_{1+x}^{\alpha} \rangle e^{ipx},
\label{structurefactor}
\end{equation}
and its increasing behavior in term of system size shows the magnetic order at the
specified momentum.
We have plotted in Fig.(\ref{Mx}) the magnetization
along $x$ direction ($M_x=(1/N)\sum \langle S_i^{x}\rangle$) and the $y$-component of spin structure factor at momentum $p=\pi$ versus $h$.  The value of $\Delta=0.25$ has been fixed to
fit the case of real material $Cs_2CoCl_4$.
The numerical data has been obtained by zero temperature Lanczos method on a finite
chain of length $N=20$ with periodic boundary condition.
\vspace{1 cm}

The magnetization curve can be distinguished in three parts:
(1) Magnetization for $0<h<h_f$,
(2) Magnetization for $h_f<h<h_c$,
and (3) the paramagnetic phase for $h>h_c$.
At zero magnetic field, there is no order in the system and all
order parameters are zero i.e $M_{x,y,z}=0$. The nonzero value of $h$ starts to align the
spins in $x$ direction  and induces a small ferromagnetic order
in $x$ direction. The magnetization, $M_x$, increases monotonically by
increasing the magnetic field. 
For $h<h_f$ the quantum effects are considerable and the magnetization changes parabolically versus $h$. 
Increasing $h$, suppresses the quantum correlations and they goes to zero around $h_f$.
For $h_f<h<h_c$, the magnetization is increased linearly versus $h$. 
Magnetization increases up to
the critical point ($h_c$) and  saturates at infinite field. Although the
staggered magnetization
along  y-direction becomes zero at the finite critical field, the magnetization in x-direction
will be fully saturated only for the isotropic case ($\Delta=1$) \cite{Reza09}.
In other words, the full saturation in
x-direction for $\Delta\neq1$ will happen for $h\rightarrow \infty$. For $h>h_c$
all of spins align almost completely (for $\Delta\neq1$) in the $x$ direction and we have a fully
polarized paramagnetic phase.

We would like to draw your attentions
to the region $h_f<h<h_c$, where the model behaves surprisingly.
Let us first study the spin chain through the entanglement ($\tau_2$) of two spins \cite{Coff00, Amic04}
which have no classical counterpart. The entanglement is defined by the following relation,
\begin{equation}
 \tau_2=\sum_{i\neq j} C_{ij}^2 \;.
\label{estimator}
\end{equation}
Where, $C_{ij}$ is the
concurrence \cite{Woot98} which is implemented instead of the
pairwise zero temperature entanglement between two spins at sites $i$ and $j$.
For $XXZ$ model, $M_z$ is zero thus the concurrence takes the
form \cite{Amic04},
$C_{ij}=2~max \{0,~ C_{ij}^{(1)},~ C_{ij}^{(2)}\},$
where
\begin{eqnarray}
\nonumber C_{ij}^{(1)}&=&\langle S_i^x S_j^x\rangle+|\langle S_i^y S_j^y\rangle-\langle S_i^z S_j^z\rangle|-\frac{1}{4},\\
\nonumber C_{ij}^{(2)}&=&|\langle S_i^y S_j^y\rangle+\langle S_i^z
S_j^z\rangle|-\sqrt{(\frac{1}{4}+\langle S_i^x
S_j^x\rangle)^2-(M^x)^2}.\\
\label{concurrence}
\end{eqnarray}

\begin{figure}[h]
\hspace{4 cm}
\includegraphics[width=10cm]{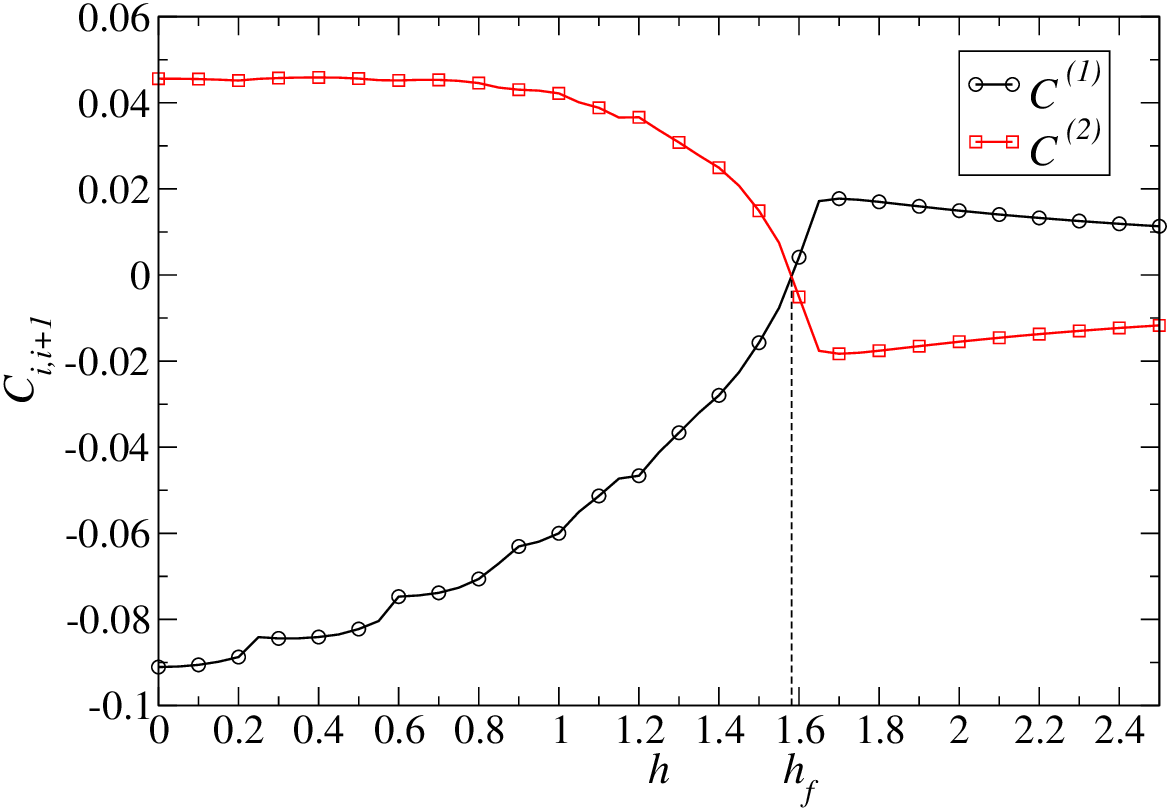}
\caption{\label{e-s} (color online) Entanglement estimators of the XXZ spin-1/2
chain versus transverse field $h$ with parameter $\Delta=0.25$. $C^{(1)}$ is depicted by (black) circles and
$C^{(2)}$ by (red) squares. At the classical field ($h=h_f\simeq1.58$),
$C_{i,i+1}^{(1)}=C_{i,i+1}^{(2)}=0$.}
\end{figure}

Using quantum Monte Carlo simulation, T. Roscilde and collaborators \cite{Rosc04}
have shown that unlike
the standard magnetic order parameters (Fig.\ref{Mx}) the pairwise entanglement, plays an essential role at
the factorizing field. At the factorizing field (also called classical field,
$h_{cl}=h_f=\sqrt{2(1+\Delta)}$) \cite{Kurm82} the ground
state takes a product form \cite{Kurm82, Reza09} and its entanglement is zero.
In Fig.(\ref{e-s}), we have plotted $C_{i,i+1}^{(1)}$
and $C_{i,i+1}^{(2)}$ versus the transverse field $h$ for XXZ spin-1/2 chain with $\Delta=0.25$ by using exact
diagonalization Lanczos method. At the classical field ($h=h_f\simeq1.58$),
$C_{i,i+1}=C_{i,i+1}^{(1)}=C_{i,i+1}^{(2)}=0$.
At this point the ground state of
the model is factorized (disentangled) where the Neel order is roughly maximized
in $y$ direction.
\vspace{1 cm}

Let us describe the behaviors of $C_{ij}^{(1)}$ and $C_{ij}^{(2)}$ from a spontaneous
symmetry breaking (SSB) point of view.
It is found in Ref. \cite{Rosc04}
that the competition between the two functions $C_{ij}^{(1)}$ and
$C_{ij}^{(2)}$ demonstrates the appearance of a SSB. In other words,
for our model, a SSB can be occurred when $C_{ij}^{(2)}<C_{ij}^{(1)}$. At
this condition the magnetic field is greater than the factorizing
value ($h>h_f$). Moreover, SSB is usually recognized to happen at
the position where the order parameter becomes zero. In our model
two standard order parameters, $M_x$ and $SM_y$ (staggered
magnetization along $y$ direction) can represent the quantum phases
of the model. We have plotted in Fig.(\ref{magnet}) the magnetization
along $x$ direction ($M_x$) and the $y$-component of spin structure
factor ($G^{yy}$) at momentum $\pi$ for $\Delta=0.25$ XXZ spin-1/2 chain. $M_x$ is nonzero in the whole range of magnetic fields
however, $G^{yy}(\pi)$ which is used to show
the antiferromagnetic order of the system has a different behavior. As it is observed from our Lanczos data,
$G^{yy}(\pi)$  increases by $h$ up to $h_f$. This behavior is also
found in the $SM^y$ curve which has been obtained by density matrix
renormalization group (DMRG) \cite{caux}.
Thus, no SSB occurs for $h<h_f$. By increasing the magnetic field
for $h>h_f$, $G^{yy}(\pi)$ (or equivalently $SM_y$) decreases
rapidly and falls to zero at the critical point $h_c$. In this
respect, a symmetry breaking can occur only for $h>h_f$.
In this region, as mentioned before the slop of $M_x$ with respect
to $h$ is different from the corresponding one for $h<h_f$.

\section{Spin waves theory \label{spwt}}

As discussed in the previous section the entanglement is zero at the factorizing field ($h_f$).
It allow
s us to write the many body ground state as the direct product of the single spin
states on odd and even sublattices \cite{Kurm82}
\begin{equation}
\label{fs}
|GS\rangle=\bigotimes_{i\in odd, j\in even}|S_{i}\rangle
|S_{j}\rangle
\end{equation}
The spin state on each sublattice is expressed in terms of polar angles ($\theta, \phi$)
which defines the rotation of the up-spin eigen-state of $S^z$ to the specific direction
defined by the factorized state. Let us label the odd (even) sublattice by $A (B)$. Thus,
($\theta, \phi$) represents A-sublattice while ($\beta, \alpha$) is the corresponding one
for B-sublattice. It has been shown \cite{Reza09} that one can consider $\phi=0=\alpha$
without loss of generality. Moreover, the magnitude of remaining polar angles
are equal $|\theta|=|\beta|$ in the case of homogeneous spin model (like here) and is given by
\begin{equation}
\cos(\theta)= -\sqrt{\frac{1+\Delta}{2}}
\end{equation}
For the antiferromagnetic model the factorized ground state is defined by $\beta=-\theta$.

Before starting the spin wave approach we implement a unitary transformation on the Hamiltonian.
All spins on the A-sublattice are rotated with angle $\theta$ counterclockwise around y-direction
and clockwise for spins on B-sublattice.
The rotated Hamiltonian ($\tilde{H}$) is the result of
rotations on all lattice points,
$\tilde{H}= \tilde{D}^{\dag}H\tilde{D}$ and
$\tilde{D}=\bigotimes_{i\in A, j\in B}
{\cal D}_i(-\theta) {\cal D}_j(\theta)$. The single spin rotation operator
is ${\cal D}_i(\theta)=exp(-i \theta S_i^y / \hbar)$.
In the rotated spin model, the Hamiltonian will have
a factorized ground state which is a ferromagnet at the factorizing field.
The rotated spin Hamiltonian is written in terms of boson operators $a, b$ with the
following Holstein-Primakoff transformation,
\begin{eqnarray}
\nonumber\tilde{S}_{Ai}^+&=&(2S-a_i^{\dagger}a_i)^{1/2}a_i,~~~~~~~~~
\tilde{S}_{Ai}^x=S-a_i^{\dagger}a_i,\\
\nonumber\tilde{S}_{Bj}^+&=&(2S-b_j^{\dagger}b_j)^{1/2}b_j,~~~~~~~~~
\tilde{S}_{Bj}^x=S-b_j^{\dagger}b_j,
\end{eqnarray}
where $\tilde{S}_{A(B)}={\cal D}_{A(B)}^{\dagger} S_{A(B)} {\cal D}_{A(B)}$ are the rotated spin operators
and ${\cal D}_{A(B)}$ is the unitary single spin rotation operator.
The bosonic Hamiltonian in the linear spin wave approximation,
i.e. $\tilde{S}_{Ai}^+\simeq\sqrt{2S} a_i; \tilde{S}_{Bi}^+\simeq\sqrt{2S} b_i$, is given by
\begin{eqnarray}
\nonumber&&\tilde{{\cal H}}=N[\frac{\Delta}{2}+\frac{h}{h_f}(1+\Delta)]+
\sum_{l=0}^N\bigg([\frac{h}{h_f}(1+\Delta)-\Delta](a_l^{\dagger}a_l+b_l^{\dagger}b_l)\\
&&-\frac{\Delta}{2}[a_l(b_l^{\dagger}+b_{l+1}^{\dagger})+h.c]
+\frac{1}{2}(\frac{h}{h_f}-1)\sqrt{1-\Delta^2}(a_l+b_l+h.c)\bigg).
\label{Hswt}
\end{eqnarray}
To diagonalize the bosonic model we first implement the Fourier transformation and then
apply a rotation to the boson operators ($a_k, b_k$)
\begin{eqnarray}
\nonumber&&a_l=\frac{1}{\sqrt{N}}\sum_k e^{-i kl}a_k,\\
\nonumber&&b_l=\frac{1}{\sqrt{N}}\sum_k e^{-i (kl+\frac{k}{2})}b_k,
\end{eqnarray}
\begin{eqnarray}
\nonumber \psi_k&=&\cos\eta_ka_k-\sin\eta_kb_k,\\
\chi_k&=&\sin\eta_ka_k+\cos\eta_kb_k.
\end{eqnarray}
The diagonalized Hamiltonian in terms of two sets of quasi-particle operators is given by
\begin{equation}
\tilde{{\cal H}}=E_0+\sum_k(\omega_k^+\chi_k^{\dagger}\chi_k+\omega_k^-\psi_k^{\dagger}\psi_k),
\label{diagonalH}
\end{equation}
where the excitation spectrums have the following forms
\begin{eqnarray}
\nonumber&&\omega_k^{\pm}=\frac{h}{h_f}(1+\Delta)-\Delta\pm\Delta\cos\frac{k}{2},\\
\nonumber&&E_0=\frac{N}{2}(\Delta-hh_f)+\omega_0^+(t^+)^2
+\sqrt{2N(1-\Delta^2)}(\frac{h}{h_f}-1)t^+,\\
&&t^+=\frac{\sqrt{\frac{N}{2}(1-\Delta^2)}(1-\frac{h}{h_f})}{\omega_0^+}.
\end{eqnarray}
In the above calculations, a translation $\chi_0\rightarrow\chi_0+t^+$ has been
performed in the diagonalization procedure of the Hamiltonian (\ref{Hswt}).

The Hamiltonian is now represented in terms of two quasi-particles (bosons),
each defines an energy scale. The two energy scales which are the excitation
energies of each bosons lead to two different dynamics for the system.
The consequence of two types of dynamics will be seen in the structure of
specific heat which will be studied. However, the existence of two different quasi-particle energies
($\omega_k^{+}\neq\omega_k^{-}$) is the first sign of two dynamics
in the model.

\subsubsection{Specific heat}

\begin{figure}[h]
\hspace{4 cm}
\includegraphics[width=10cm]{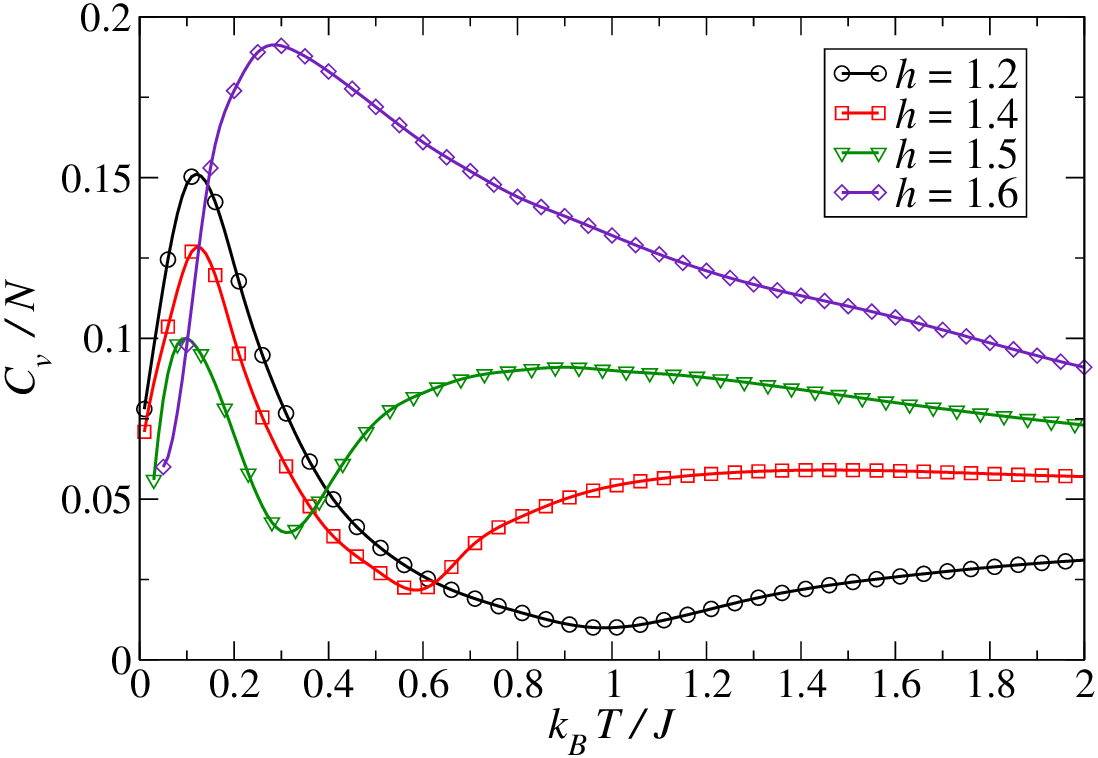}
\caption{\label{c-swt} (color online)
The spin wave results for specific heat of the XXZ model versus $T$ for
different values of the transverse magnetic field $h$ and anisotropy parameter $\Delta=0.25$. (Black) circle is for $h=1.2$, (red) square: $h=1.4$,
(green) gradient: $h=1.5$ and (blue) diamond: $h=1.6$. A narrow peak at low temperature followed
by a broaden one for higher temperature are observed for
$h=1.4, 1.5$.}
\end{figure}

To get the finite temperature properties of the model,
we assume  $\tilde{n}_k^{\pm}=\sum_{n^-,n^+}n_k^{\pm}P_k(n^+,n^-)$,
for the spin wave distribution functions, where $P_k(n^+, n^-)$ is the probability of
parallel ($n_k^+=\chi^{\dagger}_k\chi_k$) and perpendicular ($n_k^-=\psi_k^{\dagger}\psi_k$)
normal modes appearing in the $k$-momentum state which satisfies
$\sum_{n^+, n^-}P_k(n^+, n^-)=1$ for all $k$'s.
The substitutions of $\tilde{n}^+_k=\langle\chi^{\dagger}_k\chi_k\rangle$ and
$\tilde{n}_k^-=\langle\psi_k^{\dagger}\psi_k\rangle$ (where $\langle \cdots \rangle$ represents
the thermal average) in the spin-wave Hamiltonian
(\ref{diagonalH}) gives the free energy,
\begin{eqnarray}
\nonumber F&=&E_0+\sum_k(\omega_k^+\tilde{n}_k^++\omega_k^-\tilde{n}_k^-)+T\sum_k\sum_{n^+,n^-}P_k(n^-, n^+)\ln P_k(n^-, n^+).
\end{eqnarray}
The number of bosons are controlled by the following constraint which
is the magnetization in x-direction,
\begin{equation}
M_x=s-\frac{1}{2N}\sum_k(\tilde{n}_k^++\tilde{n}_k^-)-\frac{(t^+)^2}{2N}.
\label{cons}
\end{equation}
The free energy is minimized with respect to $P_k(n^+, n^-)$s under the constraint of (\ref{cons})
which is applied by a Lagrange multiplier ($\mu$) via the boson's occupation number
\begin{equation}
\tilde{n}_k^{\pm}=\frac{1}{e^{\frac{1}{k_BT}(\omega^{\pm}-\mu)}-1}.
\end{equation}
The constraint (\ref{cons}) is applied by the values of $M_x(h,T)$ which have been obtained by the numerical Lanczos method. We have plotted in Fig.(\ref{c-swt}) the specific heat of $\Delta=0.25$ XXZ spin-1/2 chain versus temperature and
different values of magnetic field. A double peak
is observed in the specific heat which is the signature of the
existence of two comparable energy scales.
More precisely, the specific heat for $h=1.4\; \mbox{and}\; 1.5$ have a narrow peak at low temperature
($T\simeq 0.2$) and a broaden one at higher $T$.
We will discuss
more on this point in the next section.
\vspace{0.5 cm}

\subsubsection{Thermal entanglement}

\begin{figure}[h]
\hspace{4 cm}
\includegraphics[width=10cm]{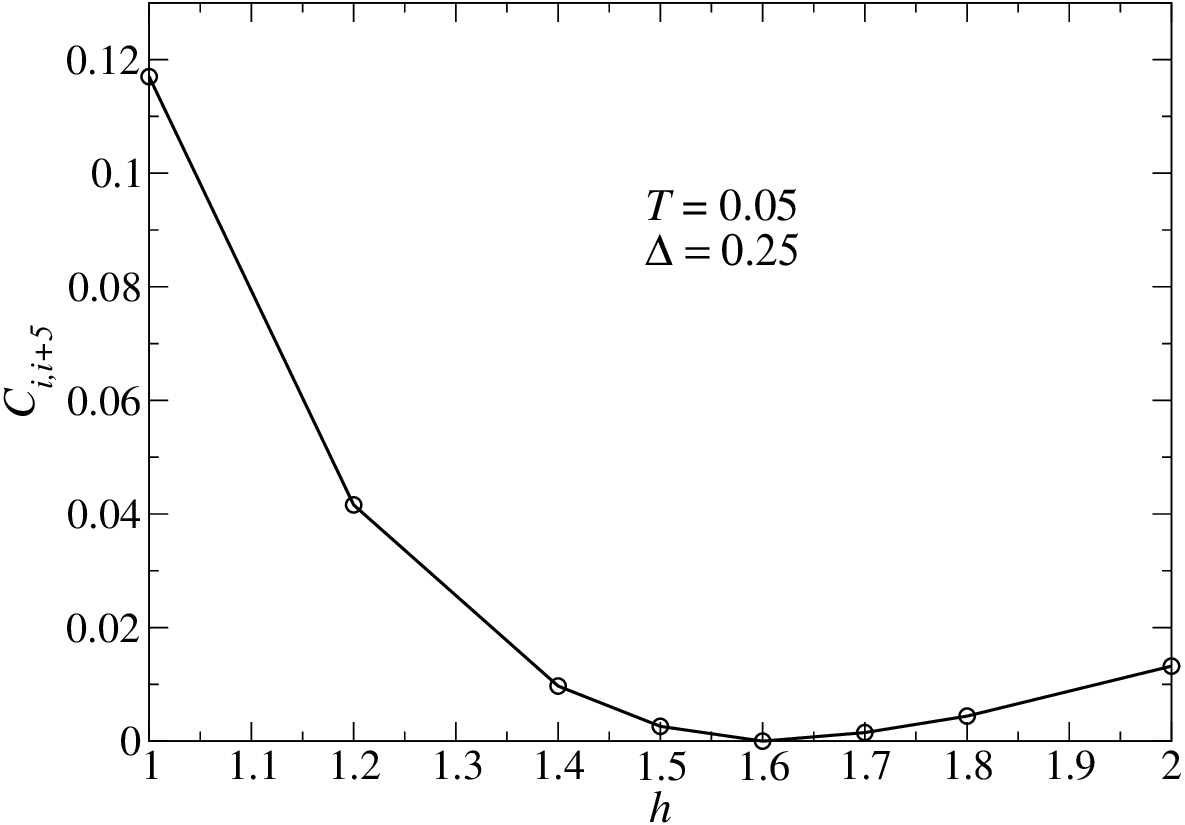}
\caption{\label{CijS} Thermal entanglement of the XXZ model versus $h$ and anisotropy parameter $\Delta=0.25$ at $T=0.05$
which has been obtained by spin wave approximation. Because of the tiny contribution of excited states to the entanglement, its value at $h_f$ is $\sim 10^{-3}$.}
\end{figure}

The established spin wave theory close to the factorizing point is used to obtain the thermal behaviors of
the correlation functions in Eqs.(\ref{concurrence}) and consequently to
calculate the concurrence. This method can describe the thermal entanglement of two spins
in linear spin wave approximation.
In this approach, one can find the following expression for $C_{ij}^{(1)}$ and $C_{ij}^{(2)}$
\begin{eqnarray}
\nonumber&&C_{ij}^{(1)}\simeq-\frac{1}{2N}\bigg(\sum_k(\tilde{n}_k^++\tilde{n}_k^-)+(t^+)^2\bigg),\\
\nonumber&&C_{ij}^{(2)}\simeq \frac{1}{2N}\bigg(\sum_k Cos(k\cdot r+k/2)(\tilde{n}_k^+-\tilde{n}_k^-)+(t^+)^2\bigg).
\end{eqnarray}
As it is seen from these functions, $C_{ij}^{(1)}$ is always less than zero thus the concurrence is $2~max\{0,C_{ij}^{(2)}\}$.
We have plotted in Fig.(\ref{CijS}) the thermal entanglement of the $\Delta=0.25$ XXZ spin-1/2 chain versus transverse field at temperature $T=0.05$. By increasing the magnetic field, quantum fluctuations are decreased and the thermal entanglement is declined. At the factorizing point, quantum fluctuations become approximately uncorrelated and the thermal entanglement is
very close to zero ($\sim 10^{-3}$). It means that the thermal entanglement is mainly
originated from the ground state and the excited states have very tiny contribution
to the entanglement.
\vspace{1 cm}

\section{Summary, Discussions and Lanczos results \label{lanczos}}

We have studied the effects of transverse magnetic field on the zero and finite temperature properties of $\Delta=0.25$ XXZ spin-1/2 chain. We have focused our attentions on the intermediate region of $h_f<h<h_c$ where the model behaves more interestingly.

M. Kenzelmann and his collaborators
\cite{Kenz02} have investigated  experimentally the effects of a transverse magnetic
field on the quasi-one dimensional spin-1/2 antiferromagnet
$Cs_2CoCl_4$, using single-crystal neutron diffraction.
Due to the weak inter-chain couplings in $Cs_2CoCl_4$
($J^{\prime}/J=0.014$) \cite{Radu 02}  where $J$ is the coupling within a chain,
it is proposed in Ref. \cite{Kenz02} that
the bulk material has a spin liquid phase at the interval $h_f<h<h_c$. In a spin liquid phase all order parameters should be zero and there should be no long range order in the system.
The existence of very weak coupling between magnetic chains in this material makes
it feasible to be described by a one dimensional spin-1/2 XXZ model with the anisotropy
parameter $\Delta=0.25$. However, for the 1D XXZ model in the region $h_f<h<h_c$ both magnetization and staggered magnetization are nonzero. Thus the ground state of the model could not be a spin liquid phase. Although the behaviors of the magnetic orders for this region are clearly known, the lacuna of a perfect study of the properties of the system at the intermediate region is still felt. In this respect, we have devoted our
attentions to survey the magnetic and thermodynamic behavior of the system at the intermediate region of the transverse field.

\begin{figure}[h]
\hspace{4 cm}
\includegraphics[width=10cm]{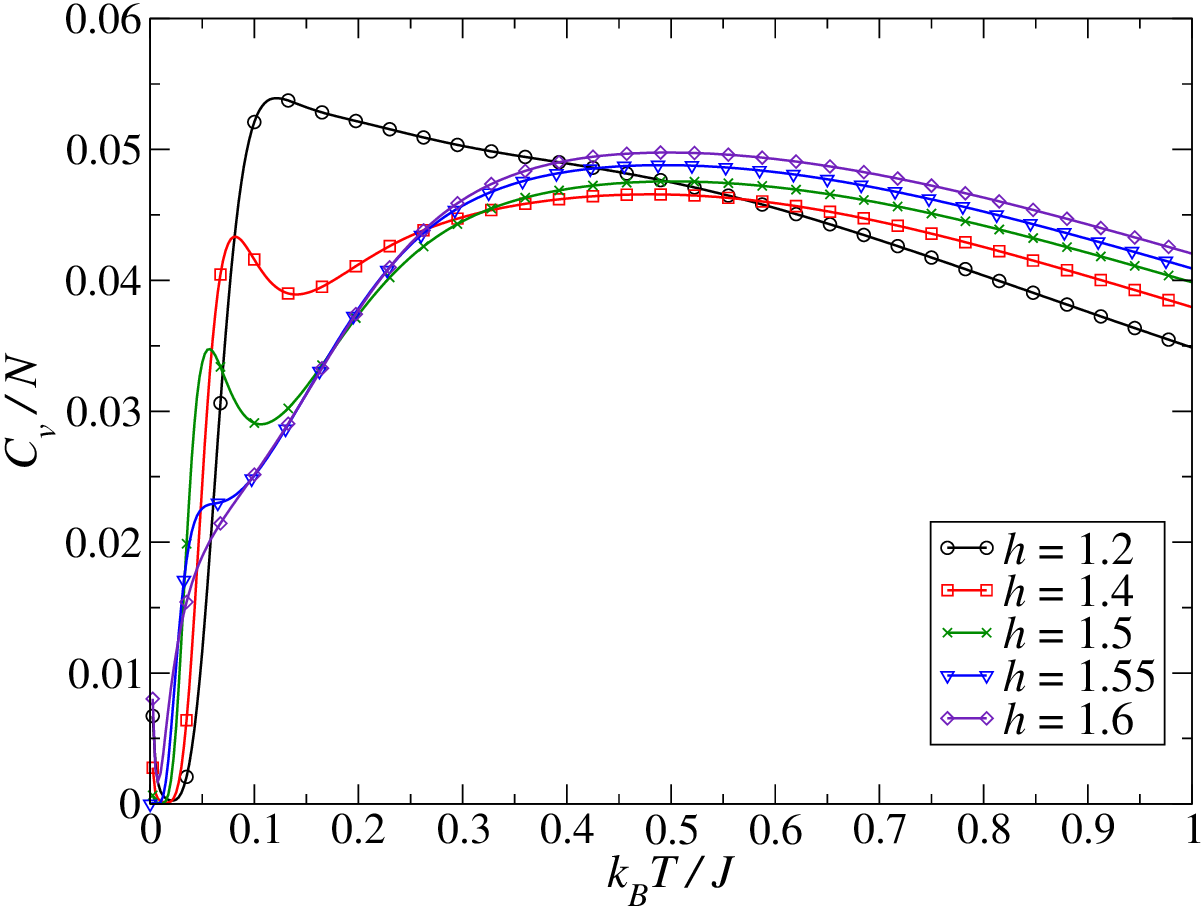}
\caption{\label{cv} The numerical Lanczos results for the specific heat of XXZ model versus $T$ for
different values of transverse magnetic field and $\Delta=0.25$. At the intermediate region of $h$, a double peak
structure is observed which indicates the existence of two energy scales in the system.}
\end{figure}

We have implemented the low temperature Lanczos method (LTLM) \cite{aichhorn,Siah08}
to compute the thermodynamic behaviors of the model for a chain of finite length.
LTLM has been used since it is accurate for low temperatures and specially the
thermodynamic averages reach the ground state expectation values as temperature
approaches zero. The specific heat versus temperature for different values of the
magnetic field on a chain of $N=20$ and with $\Delta=0.25$ has been plotted in Fig. (\ref{cv}).
The sampling is taken over $R=100$ Lanczos numerical data which have been obtained
by different initial random states. The finite size effect can be ignored since the number
of sampling ($R$) is rather high. The result of the spin wave theory (Fig.\ref{c-swt})
and LTLM (Fig.\ref{cv}) are in mutual agreement; moreover, both figures show
the presence of double peak in the specific heat for $h=1.4, 1.5$ which is
an evidence for the existence of two scales of energy or equivalently two dynamics in the system.
\vspace{1 cm}

The specific heat versus $T$ for all values of the magnetic fields has Schottky like peak
at low temperatures which is a remarkable feature of the antiferromagnetic behavior.
For small values of the magnetic field ($h<1.4$) the Schottky anomaly is the only
one which justifies the existence of a single dynamics in the model. Further increasing of the magnetic
field (being close to the factorizing point) a broaden peak is emerged in the
specific heat data versus $T$. This bump is the result of the paramagnetic order.
Two different orders are in competition and its signature is
specially observable for magnetic fields in the region $h_f\lesssim h <h_c$ where
the magnitude of the two types of ordering becomes comparable.
The antiferromagnetic order in y-direction is the result of exchange coupling
in the broken U(1) symmetry phase. The effect of transverse magnetic field as a paramagnetic order
in x-direction shows its presence when it is enough large to define
a new scale of energy. Around the factorizing field the two orders manifest their
influence on the model as a narrow and broaden peak of the specific heat which
happens for  $h_f\lesssim h <h_c$.
This is the region where the spontaneous symmetry breaking is started to happen.
At the critical point ($h_c$) the antiferromagnetic order vanishes and paramagnetic
order is the only representative of the model. A broaden peak in the specific heat versus $T$
is significant for $h>h_c$ which justifies the paramagnetic order.

Employing the specific heat data, one can also scan the behavior of the energy gap ($E_g$).
The existence of energy gap in the model is clearly observed by the exponential decay
of the specific heat at very low temperatures ($T\lesssim 0.02$).
\vspace{1 cm}
\begin{figure}
\vspace{10mm}
\centerline{\includegraphics[width=10cm]{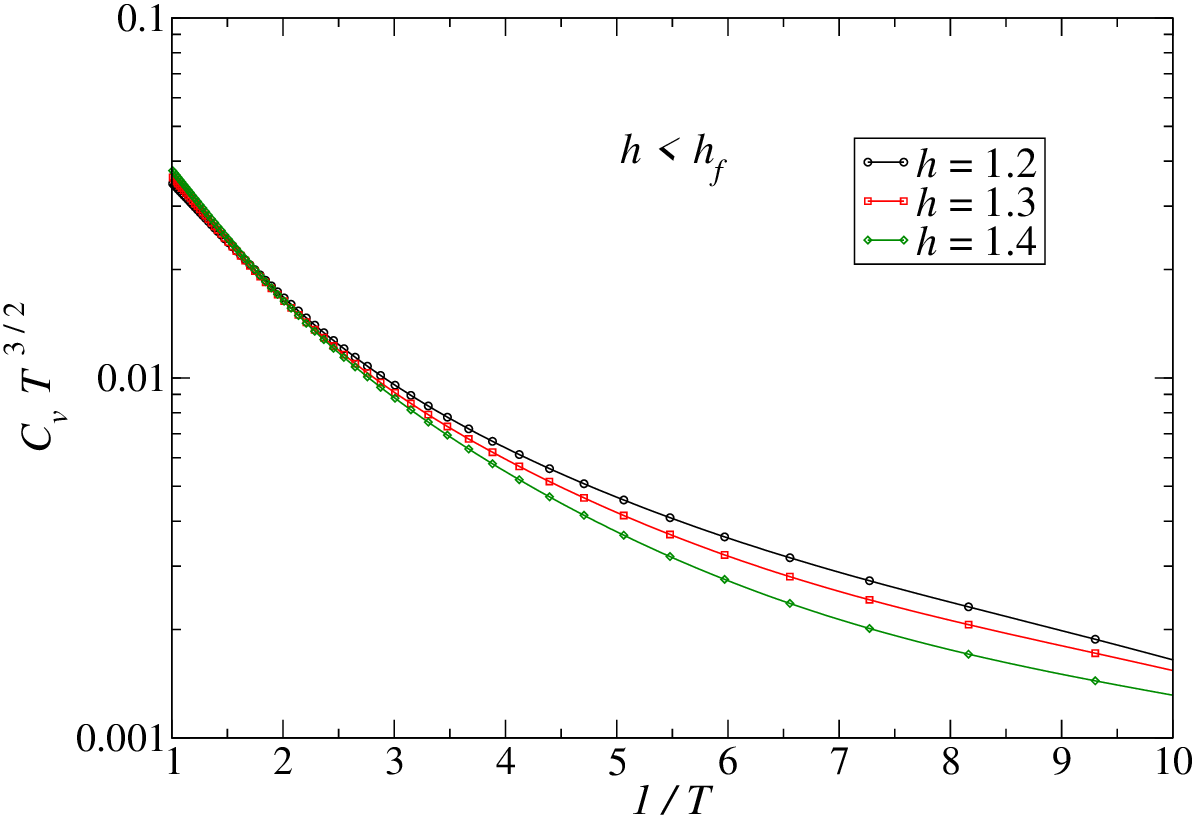}}
\caption{\label{cTb} $c_vT^{3/2}$ of XXZ model versus $1/T$
for different values of $h< h_f$ and $\Delta=0.25$. At low temperatures the slope of the curve shows increasing of the energy gap in the system.}
\vspace{10mm}
\centerline{\includegraphics[width=10cm]{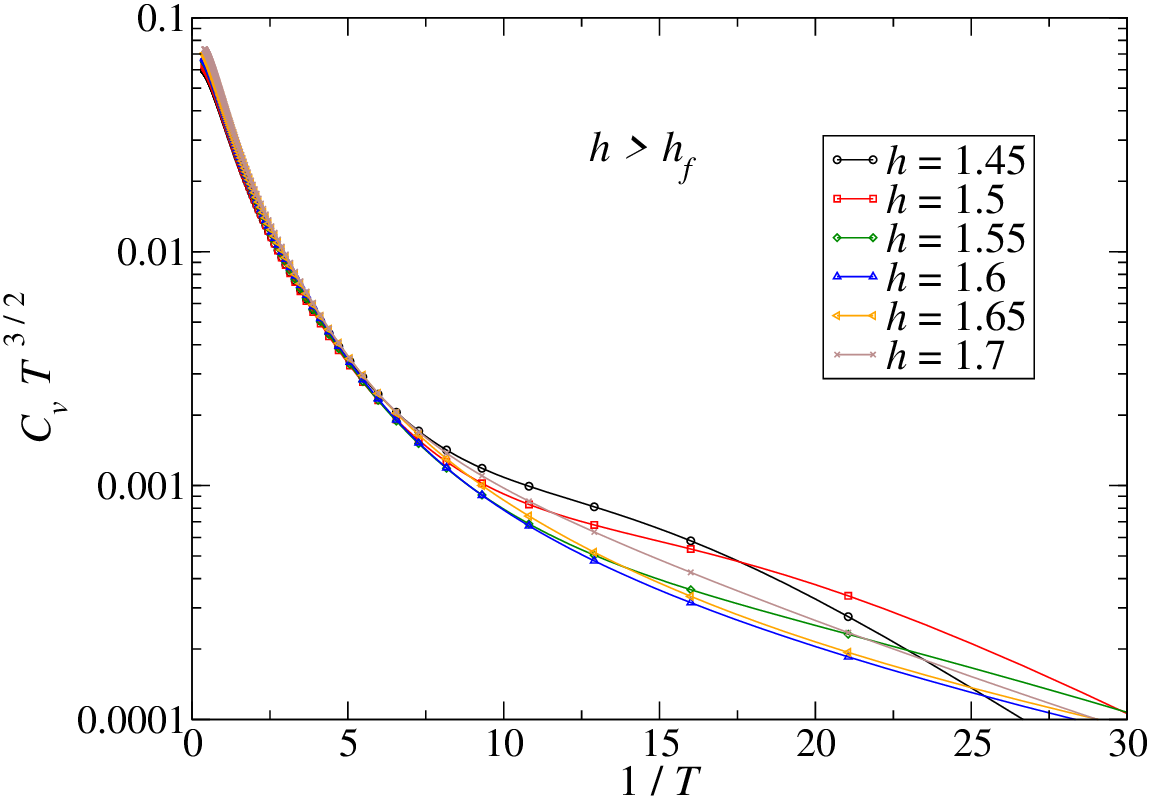}}
\caption{\label{cTa} $c_vT^{3/2}$ of XXZ model versus $1/T$
for different values of $h>h_f$ and $\Delta=0.25$. At low temperatures, the slop of the curve is decreased from $h=1.45$ to $h=1.6$ and then starts to increase by increasing of $h$. }
\end{figure}
At enough low temperatures ($T<E_g$) the specific heat and energy gap are related by \cite{Radu 02}
\begin{equation}
c_vT^{3/2}\propto e^{-\frac{E_g}{T}}.
\end{equation}
Thus, the slop of ($c_vT^{3/2}$) curves versus $\frac{1}{T}$ in log scale gives good information about the energy gap.
In Figs.(\ref{cTb},\ref{cTa}) we have plotted  ($c_vT^{3/2}$) versus $\frac{1}{T}$ for different values of transverse field.
Fig.(\ref{cTb}) shows the gap behavior close to the factorizing point, $h<h_f$.
While its behavior for $h>h_f$ is presented in Fig.(\ref{cTa}). A turning point of
different plots in Fig.(\ref{cTa}) for different transverse magnetic field is the
result of different behavior for $h<h_c\simeq 1.65$ (decreasing gap) and
$h>h_c$ (linear increasing paramagnetic gap).
Our results fit very well with
Figs. (5.14 and 5.15) of Ref. \cite{Radu 02}.

Motivated by neutron-scattering results \cite{Kenz02}, T. Radu \cite{Radu 02}
investigated the effects of non-commuting field on the ground
state of $ Cs_2CoCl_4$.
To be more precise in comparison, let us refer also to Fig.13 of chapter.5
of Ref. \cite{Radu 02} where the experimental data of the specific
heat have been shown.
The experimental data (specially in the inset of Fig.13 of chapter.5
of Ref.\cite{Radu 02}) display
a double peak in the specific heat versus temperature which is
for the intermediate range of magnetic field (less than the critical one).

It is also worth to point out the behaviors of the system from the internal energy points of view. The internal energy and specific heat are related by relation $C=\frac{dU}{dT}$. The double peak structure of the specific heat presages that at the intermediate values of the transverse field, there are two temperatures where the internal energy of the system gets its maximum variation. In other words, at these temperatures the maximum amount of states contribute to the response functions of the systems. Thus, one can also conclude that the appearance of the two energy scales 
in the system is appropriate with the number of contributed states.
Knowledge on these properties could be important in the study of the magneto-caloric effects of the XXZ model in the transverse field which is a work in progress.

\ack
We would like to thank M. Rezai for his helpful comments.
This work was supported in part by the Center of Excellence in
Complex Systems and Condensed Matter (www.cscm.ir).

\end{document}